\documentclass[12pt]{article}
\usepackage{amsfonts}
\usepackage{amssymb}
\begin{document}
\date{}
\title{\textbf{Deformed Symmetry in Snyder Space and Relativistic Particle Dynamics}}
\author{{Rabin Banerjee}\thanks{E-mail: rabin@bose.res.in}, \ {Shailesh Kulkarni}\thanks{E-mail: shailesh@bose.res.in} \ and {Saurav Samanta}\thanks{E-mail: saurav@bose.res.in}\\
\\\textit{S.~N.~Bose National Centre for Basic Sciences,}
\\\textit{JD Block, Sector III, Salt Lake, Kolkata-700098, India}}
\maketitle
                                                                                
\begin{quotation}
\noindent \normalsize 
We describe the deformed Poincare-conformal symmetries implying the 
covariance of the noncommutative space obeying Snyder's algebra.
Relativistic particle models invariant under these deformed symmetries
are presented. A gauge (reparametrisation) independent derivation of
 Snyder's algebra from such models is given. The algebraic 
transformations relating the deformed symmetries with the usual (undeformed) ones
are provided. Finally, an alternative form of an action yielding Snyder's algebra is discussed where the mass of a relativistic particle gets identified with the inverse of the noncommutativity parameter.

\end{quotation}

\section{Introduction}
Considerations based on quantum gravity and black hole physics strongly
indicate that, at the smallest scale, space-time coordinates become
 noncommutative\cite{dop}. In general the commutator among the coordinates is written as\footnote{Operators are denoted by hats to distinguish them from their classical analogues. Moreover, phase space coordinates in noncommutative space-time
 are denoted by $(\hat y, \hat q)$ in contrast to the commutative space
description given by $(\hat x, \hat p)$}
\begin{equation}
[\hat y_{\mu},\hat y_{\nu}] = i\theta_{\mu\nu}(\hat y,\hat q)\label{1.1} 
\end{equation}     
where the measure of noncommutativity $\theta_{\mu\nu}$ is taken to be a function of the phase space variables. String theory also supports relations like
(\ref{1.1})\cite{chu}. 

      There are some crucial issues related with the application of (\ref{1.1})
to physical models. In standard relativistic theory a non vanishing $\theta_{\mu\nu}$ can and does break Poincare symmetries\cite{car,ban}. Likewise for massless models, conformal symmetries are affected. However it might be possible to introduce quantum deformations of these symmetries such that the particular form of the commutator (\ref{1.1}) remains covariant. This has been discussed in great details,
for a constant $\theta_{\mu\nu}$, using either higher order differential operators\cite{wess,as,koch,rabin,lee} or twist functions following from quantum group arguments 
\cite{chai,cha}. For Lie algebraic and quadratic deformations in (\ref{1.1}), such an
analysis was done in \cite{luk}.

   Once the deformed symmetries have been defined the next issue 
 concerns the formulation of models with noncommutative space-time coordinates
 invariant under these deformations. One can of course provide a construction
 of such models but, lacking a definite prescription, these would be feasible
 in only the simplest cases, like a constant $\theta_{\mu\nu}$\cite{deri,wor}.

      In this paper we present an approach to the above problems where the noncommutativity in (\ref{1.1}) is governed by the Snyder algebra\cite{sny},
 instead of constant $\theta_{\mu\nu}$. An algebraic approach, quite distinct
 from using either the higher order derivatives or twist functions, is developed. The complete deformed conformal Poincare transformations are obtained. The 
 generators yielding these deformed transformations are found. Although the 
 generators are deformed, they satisfy the usual algebra. An explicit algebraic mapping connecting the deformed with the usual (undeformed) transformations 
 is derived. We also construct relativistic particle models that are
 invariant under the deformed symmetries.

     In section 2 we introduce the Snyder space. The usual discrete (P,T) 
symmetries are shown to be satisfied. Next, the Poincare symmetries are 
considered. A deformed translation symmetry is necessary for preserving
covariance of the Snyder algebra. Based on this analysis, a dynamical model
invariant under the deformation is formulated in section 3. This model has a deformed
symplectic structure. Using Dirac's\cite{dirac} constraint analysis and the symplectic\cite{fad} approach, the deformed brackets are computed. Either method shows that these brackets are just Snyder's brackets - the classical version of the commutator
 algebra. In section 4 we show that the deformed generators are also obtained
from Noether's theorem. Section 5 gives the algebraic map between the variables
satisfying the Snyder algebra and the standard commutative algebra. Both
classical and quantum aspects are dealt. Also, a momentum representation of 
the generators in Snyder space is given. Section 6 extends our analysis to the
conformal sector. An alternative form of an action, which describes a massive relativistic particle, yielding the Snyder algebra 
is discussed in section 7. We find that the mass of the particle gets identified with the inverse square root of the noncommutativity parameter. Our final remarks are given in section 8.                      
\section{The Snyder Space and its Symmetries}
The Snyder\cite{sny} algebra for the position and the momentum operators is given by
\begin{eqnarray}
\begin{array}{rcl}
\left[\hat y^{\mu},\hat y^{\nu}\right] &=& i\theta(\hat y^{\mu}\hat q^{\nu} - \hat y^{\nu}\hat q^{\mu})\\
\left[\hat y^{\mu},\hat q_{\nu}\right] &=& i(\delta^{\mu}_{\nu} +\theta \hat q^{\mu}\hat q_{\nu})\\
\left[\hat q_{\mu},\hat q_{\nu}\right] &=& 0 
\end{array}
\label{2.1}   
\end{eqnarray}
where ${\hat y_{\mu}}$ are the noncommutative coordinates and $\theta$ is measure of the noncommutativity. It was originally obtained by a dimensional descent from five dimensions and involves the angular momentum in the algebra of the non commuting coordinates. Taking the momentum operators commuting, as in the usual space, naturally leads to a deformed algebra among $\hat y-\hat q$, therefore ensuring the validity of the various Jacobi identities. It leads to a discrete space time compatible with Lorentz symmetry. Apart from its intrinsic interest this algebra has relevance in various contexts. For instance, a similar algebra is also obtained from quantum gravity in $2+1$ dimensions\cite{tooft}. There also exits a mapping between the Snyder space and $\kappa$-Minkowski space-time\cite{glik} which is frequently used in analysing doubly special relativity. 

We now study the different symmetries associated with the Snyder Algebra.
\subsection{Discrete (P, T) Symmetries}
 First the discrete symmetries are considered. Under the time reversal (T) operation, which is anti-linear, in the Wigner sense,
\begin{equation}
\hat y^0\rightarrow -\hat y^0, \ \hat y^i\rightarrow \hat y^i, \ \hat q_0\rightarrow
\hat q_0, \ \hat q_j\rightarrow -\hat q_j, \ i\rightarrow -i\nonumber
\end{equation}
the above algebra is invariant.

     Similarly, under the parity (P) transformation, which is linear,
\begin{equation}
\hat y^0\rightarrow \hat y^0, \ \hat y^i\rightarrow -\hat y^i, \ \hat q_0\rightarrow
\hat q_0, \ \hat q_j\rightarrow -\hat q_j, \ i\rightarrow i\nonumber
\end{equation}
 the above algebra again remains invariant. Thus both P and T symmetries are independently satisfied and no deformations are required. This may be compared with
the algebra for a constant $\theta_{\mu\nu}$ where these symmetries may be violated\cite{sheikh,haya,bala}. 
  
\subsection{Lorentz Symmetry}
By its very construction the algebra (\ref{2.1}) is compatible with standard Lorentz transformations,
\begin{eqnarray}
\delta \hat y_{\mu}&=& \omega_{\mu\alpha}\hat y^{\alpha}\label{2.2}\\
\delta \hat q_{\mu}&=& \omega_{\mu\alpha}\hat q^{\alpha}. \label{2.3}
\end{eqnarray} 
with $\omega_{\mu\alpha} =-\omega_{\alpha\mu}$. This is checked in the following way. Consider the variation in the first relation,
\begin{eqnarray}
\delta[\hat y^{\mu},\hat y^{\nu}]&=&[\delta \hat y^{\mu},\hat y^{\nu}] + [\hat y^{\mu},\delta \hat y^{\nu}]\nonumber\\
&=& i\theta \omega^{\mu\alpha}(\hat y_{\alpha}\hat q^{\nu} -\hat y^{\nu}\hat q_{\alpha})  
 - i\theta \omega^{\nu\alpha}(\hat y_{\alpha}\hat q^{\mu} -\hat y^{\mu}\hat q_{\alpha}).\label{2.4}  
\end{eqnarray}
 The same expression is obtained by considering the variation on 
 the r.h.s of that relation,
\begin{eqnarray}
i\theta \delta(\hat y^{\mu}\hat q^{\nu}-\hat y^{\nu}\hat q^{\mu}) &=& i\theta \omega^{\mu\alpha}(\hat y_{\alpha}\hat q^{\nu} -\hat y^{\nu}\hat q_{\alpha})  
 - i\theta \omega^{\nu\alpha}(\hat y_{\alpha}\hat q^{\mu} -\hat y^{\mu}\hat q_{\alpha}). \label{2.5} 
\end{eqnarray}

An identical treatment follows for the other two relations. This 
is sufficient to ensure consistency of the Lorentz transformations.
 Expectedly, the generator retains its primitive (undeformed) structure,
\begin{equation}
\hat J_{\mu\nu} = \hat y_{\mu}\hat q_{\nu} - \hat y_{\nu}\hat q_{\mu}\label{2.6}\end{equation} 
so that,
\begin{eqnarray}
\delta \hat y_{\mu} = \frac{i}{2}\omega^{\alpha\beta}\left[\hat J_{\alpha\beta},\hat y_{\mu}\right] = \omega_{\mu\alpha}\hat y^{\alpha}\label{2.7}
\end{eqnarray} 
 and similarly for $\hat q_{\mu}$.
\subsection{Translation Symmetry}
The explicit presence of the phase space variables in the algebra hints
at a possible deformation in the translation symmetry\footnote{This may be
compared with the constant $\theta_{\mu\nu}$ case of (\ref{1.1}) where translation symmetry is preserved but Lorentz symmetry
is broken.}.

     To begin with we take the transformation law for translation identical 
with the commutative space rule,
\begin{eqnarray}
\delta \hat y_{\mu}=a_{\mu}\label{2.9}\\
\delta \hat q_{\mu} = 0.\label{2.10}
\end{eqnarray}
One can easily check that this rule is not compatible with the first relation 
in the Snyder algebra. So we must change the transformation rule to achieve 
consistency.
     
      As a trial solution, general expressions of $\delta \hat y_{\mu}$ and $\delta \hat q_{\mu}$ which are dimensionally consistent are taken as    
\begin{eqnarray}
&&\delta \hat y_{\mu}=a_{\mu} + \alpha \theta a_{\mu}\hat q^2 + \beta \theta a_{\rho}\hat q^{\rho}\hat q_{\mu} \label{2.11} \\
&&\delta \hat q_{\mu} = 0.\label{2.12} 
\end{eqnarray}
Note that the transformation rule for $\hat q_{\mu}$ is kept undeformed since it is still commutative. 

       Consistency with the Snyder algebra fixes $\alpha=0$ and $\beta=1$. So the deformed transformation rule for the translation operator in Snyder space is given by,
\begin{eqnarray}
&&\delta \hat y_{\mu} = a_{\mu} + \theta a_{\rho}\hat q^{\rho}\hat q_{\mu}.\label{2.13}\\
&&\delta \hat q_{\mu} = 0.\label{2.14}
\end{eqnarray}

      Although we have a deformed transformation rule for translation, the generator remains the same as in the commutative space. To emphasize this point we note that 
\begin{eqnarray}
\delta \hat y^{\mu} &=& i\left[\hat G,\hat y^{\mu}\right] = ia^{\rho}[\hat q_{\rho},\hat y^{\mu}]\nonumber\\
&=& a^{\mu} + \theta a^{\rho}\hat q_{\rho}\hat q^{\mu}.\label{2.15}
\end{eqnarray}
and likewise for $\hat q_{\mu}$.
   
      Thus the Poincare generators in Snyder space and usual commutative space are form invariant. However, whereas Lorentz transformation remains undeformed,
translation get deformed. 

     Finally, in spite of the involved algebra (\ref{2.1}) these generators satisfy the usual Poincare algebra,
\begin{eqnarray}
\begin{array}{rcl}
&&[\hat q_{\mu},\hat q_{\nu}]=0\\
&&\left[\hat J_{\mu\nu},\hat q_{\lambda}\right]=i(\delta_{\mu\lambda}\hat q_{\nu}-\delta_{\nu\lambda}\hat q_{\mu})\\
&&\left[\hat J_{\mu\nu},\hat J_{\rho\sigma}\right]=-i(\delta_{\nu\rho}\hat J_{\mu\sigma}+\delta_{\mu\sigma}\hat J_{\nu\rho}-\delta_{\mu\rho}\hat J_{\nu\sigma}-\delta_{\nu\sigma}\hat J_{\mu\rho}).
\end{array}
\label{2.16}
\end{eqnarray}
\section{Dynamical Models Invariant Under Deformation and the Snyder Algebra} 
Here a nontrivial application of the deformed symmetries is provided. Specifically, we discuss a method by which dynamical models can be constructed to yield, from  their symplectic structure, the Snyder algebra. Several authors\cite{rom,gir,ghosh,rome} have suggested various models leading to this algebra but they lack a clear cut guiding principle. This is further exemplified by the fact that the results are obtained in a 
specific gauge. The Snyder algebra therefore occurs as an artefact of the gauge rather than something fundamental. 

        We adopt the following strategy. A dynamical model is constructed that
is invariant under the deformed (translation)  symmetry. The ensuing model has
an involved symplectic structure which is elucidated by both the Dirac and symplectic approaches. A calculation of the Dirac (or symplectic) brackets yields Snyder
 algebra. No gauge (or reparametrisation) fixing is necessary. It is also
reassuring to note that, following a Noether approach, we show that the 
Poincare generators remain form invariant, exactly as discussed in the previous
section.              
 
    Consider the following first order form of the action for a relativistic free particle of mass $m$, 
\begin{eqnarray}
S=\int d\tau[-\dot q^{\mu}y_{\mu}-e(q^2-m^2)]
\label{3.1}
\end{eqnarray}
where $e$ is a Lagrange multiplier enforcing the Einstein condition $q^2-m^2=0$.

  Since the Lorentz transformation is undeformed, obviously (\ref{3.1}) remains invariant. Under translation however,
\begin{eqnarray}
\delta S&=&\int d\tau[-\dot q^{\mu}(a_{\mu}+\theta a_{\rho}q^{\rho}q_{\mu})]\\
 &=&\int d\tau[-\frac{d}{d\tau}(q^{\mu}a_{\mu})- \theta a_{\rho}q^{\rho}q_{\mu}\dot q^{\mu}]\label{3.3}     
\end{eqnarray} 
obtained on exploiting (\ref{2.13}), (\ref{2.14}). The additional symmetry breaking term can be written as, 
\begin{eqnarray}
\theta a_{\rho}q^{\rho}q_{\mu}\dot q^{\mu}&=&\theta(\delta y_{\rho}-\theta a_{\sigma}q^{\sigma}q_{\rho})q^{\rho}q_{\mu}\dot q^{\mu}\label{3.4}\\
&=&\theta\delta[ y_{\rho}q^{\rho}q_{\mu}\dot q^{\mu}]-\theta^2\delta[y_{\sigma}q^{\sigma}q^2\dot q^{\mu}q_{\mu}]+\theta^3\delta[y_{\sigma}q^{\sigma}(q^2)^2\dot q^{\mu}q_{\mu}]+\cdot\cdot\cdot\label{3.5}\\
&=&\theta\delta[\frac{1}{1+\theta q^2}(y\cdot q)\dot q^{\mu}q_{\mu}]\label{3.6}
\end{eqnarray}
where recursive use of (\ref{2.13}) and (\ref{2.14}) has been done. 

  It is clear that by introducing the term inside the parentheses in the action, invariance under deformed translations would be preserved\footnote{Actually a total time derivative $\frac{d}{d\tau}(-q^{\mu}a_{\mu})$ remains but this is allowed. It is in fact related to the generator, as shown later.}. Thus the relevant action is given by,
\begin{eqnarray}
S=\int d\tau[-\dot q^{\mu}y_{\mu}+\frac{\theta}{1+\theta q^2}(y.q)\dot q^{\mu}q_{\mu}-e(q^2-m^2)].\label{3.7}
\end{eqnarray}
This is invariant under Lorentz transformations,
\begin{eqnarray}
\delta S=0
\end{eqnarray}
and quasi-invariant under deformed translations,
\begin{eqnarray}
\delta S=\int d\tau[-\frac{d}{d\tau}(a^{\mu}q_{\mu})].\label{w}
\end{eqnarray}
\subsection{Dirac's Constraint Analysis}
We interpret $y$ and $q$ of the first order action (\ref{3.7}) as the configuration variables in an extended space. The canonical momentum conjugate to $y$, $q$ and $e$ are,
\begin{eqnarray}
\pi_{\mu}^{y}&=&\frac{\partial L}{\partial \dot y^{\mu}}=0\nonumber\\
\pi_{\mu}^{q}&=&\frac{\partial L}{\partial \dot q^{\mu}}=-y_{\mu}+\frac{\theta}{1+\theta q^2}(y.q)q_{\mu}\nonumber\\
\pi_{e}&=&\frac{\partial L}{\partial \dot e}=0.\nonumber
\end{eqnarray}
Since none of the momenta involve velocities these have to be interpreted, following Dirac\cite{dirac}, as primary constraints. These are given by
\begin{eqnarray}
&&\Phi=\pi_{e}\approx0\label{3.8}\\
&&\Phi_{1,\mu}=\pi_{\mu}^{y}\approx0\label{3.9}\\
&&\Phi_{2,\mu}=\pi_{\mu}^{q}+y_{\mu}-\frac{\theta}{1+\theta q^2}(y.q)q_{\mu}\approx0.\label{3.10}
\end{eqnarray}
The Poisson algebra of the constraints is given by
\begin{eqnarray}
&&\{\Phi,\Phi\}=\{\Phi,\Phi_{1,\mu}\}=\{\Phi,\Phi_{2,\mu}\}=0\label{3.11}\\
&&\{\Phi_{1,\mu},\Phi_{1,\nu}\}=0\label{3.12}\\
&&\{\Phi_{1,\mu},\Phi_{2,\nu}\}=-\eta_{\mu\nu}+\frac{\theta}{1+\theta q^2}q_{\mu}q_{\nu}\label{3.13} \\
&&\{\Phi_{2,\mu},\Phi_{2,\nu}\}=\frac{\theta}{1+\theta q^2}(q_{\nu}y_{\mu}-q_{\mu}y_{\nu})\label{3.14}. 
\end{eqnarray}
Because the algebra of the constraints $\Phi_{1,\mu}$ and $\Phi_{2,\mu}$ does not close, they form a second class set. This set can be eliminated as shown later, by the use of Dirac brackets.

    Since the action given in (\ref{3.7}) is first order, the canonical Hamiltonian of the system can be written easily
\begin{eqnarray}
H_C=e(q^2-m^2).\label{3.15}
\end{eqnarray}
Following Dirac the total Hamiltonian is given by
\begin{eqnarray}
H_T=e(q^2-m^2)+\lambda \Phi+\lambda_{1,\mu}\Phi_{1,\mu}+\lambda_{2,\mu}\Phi_{2,\mu}.\label{3.17}
\end{eqnarray}
Time consistency of the constraint (\ref{3.8}) leads to the following secondary constraint
\begin{eqnarray}
\Psi=\{H_T,\pi_e\}=q^2-m^2\approx0.\label{3.18}
\end{eqnarray}

  The second class constraint sector $\Phi_1,\Phi_2$ is next eliminated by using Dirac brackets. The first step is to compute the constraint matrix,
\begin{eqnarray}
\Lambda_{\mu\nu}&=&\left(\matrix{\{\Phi_{1,\mu},\Phi_{1,\nu}\}&\{\Phi_{1,\mu},\Phi_{2,\nu}\}\cr\{\Phi_{2,\mu},\Phi_{1,\nu}\}&\{\Phi_{2,\mu},\Phi_{2,\nu}\}}\right)\label{3.19}\\
&=&\left(\matrix{0 & -\eta_{\mu\nu}+\frac{\theta}{1+\theta q^2}q_{\mu}q_{\nu} \cr \eta_{\mu\nu}-\frac{\theta}{1+\theta q^2}q_{\mu}q_{\nu} & \frac{\theta}{1+\theta q^2}(q_{\nu}y_{\mu}-q_{\mu}y_{\nu}) }\right).\label{3.20}
\end{eqnarray}
We write the inverse of $\Lambda_{\mu\nu}$ as $\Lambda^{\mu\nu}$ such that $\Lambda_{ij}^{\mu\nu}\Lambda_{jk,\nu\rho}=\delta^{\mu}_{ik,\rho} \ (i,j=1,2)$.
It is given by
\begin{eqnarray}
\Lambda^{\mu\nu}=\left(\matrix{\theta(y^{\mu}q^{\nu}-y^{\nu}q^{\mu})& \eta^{\mu\nu}+\theta q^{\mu} q^{\nu}\cr -\eta^{\mu\nu}-\theta q^{\mu} q^{\nu}&0  }\right).\label{3.21}
\end{eqnarray}
At this point one can calculate the various Dirac brackets using the definition\cite{dirac}.
\begin{eqnarray}
\{f,g\}_{DB}=\{f,g\}-\{f,\Phi_{i,\mu}\}\Lambda_{ij}^{\mu\nu}\{\Phi_{j,\nu},g\}.\label{3.22}
\end{eqnarray}
For our model the Dirac brackets among the configuration space variables are
\begin{eqnarray}
\begin{array}{rcl}
&&\{y^{\mu},y^{\nu}\}_{DB}=\theta\left(y^{\mu} q^{\nu}- y^{\nu} q^{\mu}\right )\\
&&\{q_{\mu},q_{\nu}\}_{DB}=0\\
&&\{y^{\mu},q_{\nu}\}_{DB}=\delta^{\mu}_{ \nu}+\theta q^{\mu}q_{\nu}.
\end{array}
\label{3.23}
\end{eqnarray}

This algebra is basically the classical version of the Snyder algebra given in (\ref{2.1}). In order to elevate this algebra at the operator level we note that there is no ordering problem.
 Since $q$ s commutes among themselves there is no problem in the algebra between $y^{\mu}$ and $q_{\nu}$. Furthermore the difference between $y^{\mu} q^{\nu}$ and $q^{\nu}y^{\mu}$ is symmetrical in $\mu,\nu$. Since the bracket between $y^{\mu}$ and $y^{\nu}$ is antisymmetric in $\mu,\nu$ there is no ordering problem in this case also. Consequently the Dirac brackets (\ref{3.23}) get lifted to the commutators (\ref{2.1})
 without ordering ambiguities. A generalized version of Snyder algebra, where both $y$ and $q$ are noncommuting, has been discussed in \cite{jaro}.

   Since the constraint $\Phi$ is the canonical conjugate momentum of the Lagrange multiplier $e$, it is not physically important. On the other hand the secondary constraint $\Psi$ has vanishing Dirac brackets with all constraints,
\begin{eqnarray}
&&\{\Psi,\Psi\}_{DB}=\{\Psi,\Phi\}_{DB}=0\label{3.24}\\
&&\{\Psi,\Phi_{1,\mu}\}_{DB}=\{\Psi,\Phi_{2,\mu}\}_{DB}=0.\label{3.25}
\end{eqnarray}
Therefore the constraint $\Psi$ is first class and hence the generator of gauge (reparametrisation) transformation.

    It is now useful to note that $q_{\mu}$ and $J_{\mu\nu}$ are gauge invariant variables since,
\begin{eqnarray}
&&\{q_{\mu},(q^2-m^2)\}_{DB}=0\label{3.26}\\
&&\{J_{\mu\nu},(q^2-m^2)\}_{DB}=0.\label{3.27}
\end{eqnarray}
For this reason the  Dirac algebra and Poisson algebra involving $q_{\mu}$ and $J_{\mu\nu}$ are identical. Since their Poisson algebra leads to the usual 
Poincare algebra it is clear that their Dirac algebra also yields the same
Poincare algebra. It gives a dynamical explanation of the fact that $J_{\mu\nu}$ and $q_{\nu}$ satisfy the undeformed Poincare algebra inspite of the deformed
 algebra (\ref{2.1}) of its composites. The argument is equally valid at the quantum level since (\ref{3.26}) and (\ref{3.27}) may be elevated to commutators by
\begin{eqnarray}
&&[\hat q_{\mu},(\hat q^2-m^2)]=0\label{3.28}\\
&&[\hat J_{\mu\nu},(\hat q^2-m^2)]=0\label{3.29}
\end{eqnarray}
which may be easily verified by using the basic algebra.
\subsection{Symplectic Analysis}
As is well known there is an alternative (and occasionally quicker than Dirac's) method of getting the basic brackets. This is the symplectic approach\cite{fad} and is geared for first order systems. In this method everything is obtained from the equations of motion and the obtention or classification of constraints is redundant.

    The equations of motion obtained from (\ref{3.7}) are
\begin{eqnarray}
&&\dot q_{\mu}-\frac{\theta}{1+\theta q^2}(\dot q.q)q_{\mu}=0\label{4.1}\\
&&\dot y_{\mu}-\frac{\theta}{1+\theta q^2}\{(\dot y.q)q_{\mu}-(\dot q.y)q_{\mu}+(\dot q.q)y_{\mu}\}-2q_{\mu}=0. \label{4.2}
\end{eqnarray}
These equations of motion can be written in the form
\begin{eqnarray}
\Lambda_{ij,\mu\nu}\dot \xi_{j,\nu}=\frac{\partial H_C}{\partial\xi_{i,\mu}}\label{4.3}
\end{eqnarray}
where
\begin{eqnarray}
&&\xi_1^{\mu}=y^{\mu},\label{4.4}\\
&&\xi_2^{\mu}=q^{\mu}\nonumber
\end{eqnarray}
and $\Lambda_{ij,\mu\nu}$ is given in (\ref{3.20}). The canonical Hamiltonian $H_C$ is defined in (\ref{3.15}).

The symplectic brackets are given by
\begin{eqnarray}
\{f,g\}_{SB}=\Lambda_{ij}^{\alpha\beta}\partial_{i,\alpha}f\partial_{j,\beta}g\label{4.5}
\end{eqnarray}
where
\begin{eqnarray}
\partial_{i,\alpha}=
\frac{\partial}{\partial \xi_i^{\alpha}}\label{4.6}
\end{eqnarray}
and $\Lambda_{ij}^{\alpha\beta}$ is given in (\ref{3.21}).
So the relevant symplectic brackets are
\begin{eqnarray}
\begin{array}{rcl}
&&\{y^{\mu},y^{\nu}\}_{SB}=\theta\left(y^{\mu} q^{\nu}- y^{\nu} q^{\mu}\right )\\
&&\{q_{\mu},q_{\nu}\}_{SB}=0\\
&&\{y^{\mu},q_{\nu}\}_{SB}=\delta^{\mu}_{ \nu}+\theta q^{\mu}q_{\nu}.
\end{array}
\label{4.7}
\end{eqnarray}
The symplectic brackets are identical with the Dirac brackets and
 generate the classical version of the Snyder algebra given in (\ref{2.1}).
\section{Noether's Theorem and Generators}
It is possible to reproduce the Poincare generators from a Noether analysis of (\ref{3.7}). This provides a link between the algebraic way of obtaining the generators in section 2 and the dynamical method.

  In general, the invariance of an action $S$ under an infinitesimal symmetry transformation,
\begin{eqnarray}
\delta Q_i=\{Q_i,G\}\label{5.1}
\end{eqnarray}
is given by
\begin{eqnarray}
\delta S=\int d \tau \frac{d}{d\tau}(\delta Q^{\mu}P_{\mu}-G)\label{5.2}
\end{eqnarray}
where $G$ is the generator of the transformation and $P_{\mu}$ is the canonical momenta conjugate to $Q^{\mu}$. If the quantity inside the parentheses is denoted by $B(Q,P)$, then the generator is defined as,
\begin{eqnarray}
G=\delta Q^{\mu}P_{\mu}-B.\label{5.3}
\end{eqnarray}
For the model (\ref{3.7}) both $y,q$ are interpreted as configuration space variables so that,
\begin{eqnarray}
G=\delta q^{\mu}\pi_{\mu}^q+\delta y^{\mu}\pi_{\mu}^y-B.\label{5.4}
\end{eqnarray}
This is further simplified on using the constraints (\ref{3.9}), (\ref{3.10}) to yield,
\begin{eqnarray}
G=\delta q^{\mu}(\frac{\theta}{1+\theta q^2}(y\cdot q)q_{\mu}-y_{\mu})-B.\label{5.5}
\end{eqnarray}
\vspace{0.40cm}\\
{\bf Translations } \ : \vspace{0.40cm}\\
   For translations (\ref{2.14}) and (\ref{w}) reveal that,
\begin{eqnarray}
\delta q^{\mu}=0, \ B=-a^{\sigma}q_{\sigma}\label{5.6}
\end{eqnarray}
so that,
\begin{eqnarray}
G=a^{\sigma}q_{\sigma}\label{5.7}
\end{eqnarray}
yielding the cherished expression.
\vspace{0.40cm}\\
 {\bf Rotations} \ : \vspace{0.40cm}\\ 
   For rotations $\delta q^{\mu}$ is given by (\ref{2.3}) while $B=0$ since the Lagrangian itself is manifestly invariant ($\delta L=0$). Hence we get,
\begin{eqnarray}
G&=&\omega^{\mu\alpha}q_{\alpha}(\frac{\theta}{1+\theta q^2}(y\cdot q)q_{\mu}-y_{\mu})\nonumber\\
&=&-\omega^{\mu\alpha}q_{\alpha}y_{\mu}\nonumber\\
&=&\frac{\omega^{\mu\alpha}}{2}J_{\alpha\mu}\label{5.8}
\end{eqnarray}
which is the desired form of the rotation generator.
\section{Mapping Between Deformed and Usual Symmetries}
In this section we discuss algebraic transformations mapping the deformed symmetries with the primitive (undeformed) ones. This is obtained by comparing the actions (\ref{3.1}) and (\ref{3.7}). The action (\ref{3.1}) satisfies the undeformed symmetries while (\ref{3.7}) satisfies their deformed versions. Now (\ref{3.1}) is rewritten in terms of canonical variables $(x,p)$ so that,
\begin{eqnarray}
S=\int d\tau[-\dot p_{\mu}x^{\mu}-e(p^2-m^2)]\label{6.1}
\end{eqnarray}
with,
\begin{eqnarray}
\{x^{\mu},p_{\nu}\}=\delta_{\mu\nu}, \ \{x^{\mu},x^{\nu}\}=\{p_{\mu},p_{\nu}\}=0\label{6.2}
\end{eqnarray}
Also, under translations and Lorentz transformations,
\begin{eqnarray}
&&\delta x^{\mu}=a^{\mu}, \ \delta p_{\mu}=0\label{6.3}\\
&&\delta x^{\mu}=\omega^{\mu\alpha}x_{\alpha}, \ \delta p_{\mu}=\omega_{\mu\alpha}p^{\alpha}.\label{6.4}
\end{eqnarray}
Then the actions (\ref{6.1}) and (\ref{3.7}) are mapped by the transformations,
\begin{eqnarray}
&& x_{\mu}=y_{\mu}-\frac{\theta}{1+\theta q^2}(y\cdot q)q_{\mu}\label{6.5}\\
&& p_{\mu}=q_{\mu}.\label{6.6}
\end{eqnarray}
The maps preserve the stability of the infinitesimal Poincare transformations. For instance, using (\ref{2.13}) and (\ref{2.14}) we obtain for translations,
\begin{eqnarray}
\delta(y_{\mu}-\frac{\theta}{1+\theta q^2}(y\cdot q)q_{\mu})=a_{\mu}=\delta x_{\mu}\label{6.7}
\end{eqnarray}
thereby verifying our assertion. Rotations are trivially preserved.

 The inverse map is given by,
\begin{eqnarray}
&&y_{\mu}=x_{\mu}+\theta(x\cdot p)p_{\mu}\label{6.8}\\
&&q_{\mu}=p_{\mu}.\label{6.9}
\end{eqnarray}
The classical Snyder algebra follows from the above relations by using the canonical algebra (\ref{6.2}),
\begin{eqnarray}
\{y^{\mu},q_{\nu}\}&=&\{x^{\mu}+\theta(x\cdot p)p^{\mu},p_{\nu}\}\label{6.10}\\
&=&\delta^{\mu}_{\nu}+\theta q^{\mu}q_{\nu}\label{6.11}
\end{eqnarray}
and likewise for the other brackets.

  It is feasible to construct operator analogue of the maps (\ref{6.5}), (\ref{6.6}) by giving an ordering prescription. Using the Weyl (symmetric) ordering, we get,
\begin{eqnarray}
\hat x^{\mu}=\hat y^{\mu}&-&\frac{\theta}{8}[\frac{ \hat q^{\mu} \hat  q_{\rho}}{1+\theta \hat q^2}\hat y^{\rho} +  \hat q^{\mu} \hat q_{\rho} \hat  y^{\rho}\frac{1}{1+\theta  \hat q^2}\nonumber\\
&& + \frac{ \hat q^{\mu}}{1+\theta \hat q^2}\hat y^{\rho} \hat q_{\rho}
+ \frac{\hat q_{\rho}}{1+\theta  \hat q^2}\hat y^{\rho} \hat q^{\mu}\nonumber\\ 
&& +  \hat q^{\mu} \hat y^{\rho}\frac{ \hat q_{\rho}}{1+\theta  \hat q^2}
 + \hat q_{\rho}\hat y^{\rho} \frac{ \hat q^{\mu}}{1 +\theta  \hat q^2}\label{6.12}\\
&& + \frac{1}{1+\theta  \hat q^2}\hat y^{\rho}  \hat q^{\mu} \hat q_{\rho} 
+ \hat y^{\rho} \frac{ \hat q^{\mu}  \hat q_{\rho}}{1+\theta  \hat q^2}].\nonumber\\
\hat p_{\mu}=\hat q_{\mu}.\nonumber
\end{eqnarray}

The inverse transformation is found to be,
\begin{eqnarray}
\begin{array}{rcl}
&&\hat y^{\mu}=\hat x^{\mu}+\frac{\theta}{4}[\hat x^{\rho}\hat p_{\rho}\hat p^{\mu}+\hat p^{\mu}\hat p_{\rho}\hat x^{\rho}+ \hat p^{\mu}\hat x^{\rho}\hat p_{\rho}+\hat p_{\rho}\hat x^{\rho}\hat p^{\mu}]\\
&&\hat q_{\mu}=\hat p_{\mu}
\end{array}
\label{6.13}                              
\end{eqnarray}
which is just the Weyl ordered form of (\ref{6.8}), (\ref{6.9}). 

       A slightly lengthy computation reveals that the quantum Snyder algebra (\ref{2.1}) as a commutator algebra follows from (\ref{6.13}) by using the standard canonical commutators involving $x$ and $p$. 
   
       It is also possible to prove, once again after some algebra, the operator identity obtained from(\ref{6.13}), 
\begin{equation}
\hat x_{\mu}\hat p_{\nu} - \hat x_{\nu}\hat p_{\mu} =  \hat y_{\mu}\hat q_{\nu} - \hat y_{\nu}\hat q_{\mu} = \hat J_{\mu\nu}
\label{6.14}
\end{equation}  
which illustrates the form invariance of the angular momentum generator. It
also provides another explanation of the fact that the Poincare generators 
in the Snyder basis satisfy the usual algebra.\\
\vspace{0.40cm}\\
{\bf Representation} \ : \vspace{0.40cm}\\
  It is possible to give a particular representation for the operators $\hat y_{\mu},\hat q_{\nu}$. Since the momenta $(\hat q_{\mu})$ commute, momentum representation is favored. One may verify that the differential representation of the operators leading to the Snyder algebra (\ref{2.1}) is given by,
\begin{eqnarray}
&&\hat q_{\mu}= q_{\mu}\label{6.15}\\
&&\hat y_{\mu}=i\left(\frac{\partial}{\partial  q^{\mu}}+\theta  q_{\mu}q_{\nu}\frac{\partial}{\partial q_{\nu}}\right).\label{6.16}
\end{eqnarray}
This representation was also obtained in\cite{jaro}. Moreover the space part of these relations had occurred earlier in a different context\cite{chang}.
As an application of this representation, the identity (\ref{6.14}) is reproduced. The angular momentum operator is represented as,
\begin{eqnarray}
\hat J_{\mu\nu}&=&\hat y_{\mu}\hat q_{\nu}-\hat y_{\nu}\hat q_{\mu}\label{6.17}\\
&=&\hat q_{\nu}\hat y_{\mu}-\hat q_{\mu}\hat y_{\nu}\label{6.18}\\
&=&i\left(q_{\nu}\frac{\partial}{\partial q^{\mu}}-q_{\mu}\frac{\partial}{\partial q^{\nu}}\right)\label{6.19}
\end{eqnarray}
which just corresponds to the usual (momentum space) representation of angular momentum in the commutative space.

\section{Deformed Conformal Symmetry}
After discussing the translational and the Lorentz symmetry we next
consider deformations in the dilatation and the special conformal transformation.
First an algebraic analysis is considered which is followed by a dynamical treatment related to the action (\ref{3.7}).
\subsection{Dilatation Symmetry}
Let us begin by treating the usual transformations under dilatation,
\begin{eqnarray}
&&\delta \hat y_{\mu} = \epsilon \hat y_{\mu}\label{7.1}\\
&&\delta \hat q_{\mu} = - \epsilon \hat q_{\mu}.\label{7.2}   
\end{eqnarray}
It is clear that covariance of only the last relation in (\ref{2.1}) is preserved. Thus 
although the transformation for $\hat q_{\mu}$ is unmodified, that for $\hat y_{\mu}$  
must be deformed. We take as an ansatz
\begin{equation}
\delta \hat y_{\mu} = \epsilon \hat y_{\mu} + \epsilon \hat Q_{\mu}(\theta).\label{7.3}
\end{equation} 
Then demanding covariance of the second relation in (\ref{2.1}) one obtains,
\begin{eqnarray}
[\delta \hat y_{\mu},\hat q_{\nu}] + [\hat y_{\mu},\delta \hat q_{\nu}] &=& i\delta(\delta_{\mu\nu}
+ \theta \hat q_{\mu}\hat q_{\nu})\label{7.4}\\
&=& -2i\theta \epsilon \hat q_{\mu}\hat q_{\nu}\label{7.5}
\end{eqnarray}
which yields
\begin{equation}
[\hat Q_{\mu},\hat q_{\nu}] = -2i\theta \hat q_{\mu}\hat q_{\nu}.\label{7.6}
\end{equation}
Up to an ordering ambiguity a solution for $\hat Q_{\mu}$ is given by
\begin{equation}
\hat Q_{\mu} = -\frac{2\theta(\hat y\cdot \hat q)\hat q_{\mu}}{1+\theta \hat q^2}.\label{7.7}
\end{equation}
The ambiguity is fixed by requiring covariance of the $\hat y_{\mu}-\hat y_{\nu}$ bracket
in (\ref{2.1}) leading to the following transformation law,
\begin{eqnarray}
\delta \hat y_{\mu} &=& \epsilon[\hat y_{\mu}-\hat y_{\rho}\frac{\theta \hat q^{\rho}\hat q_{\mu}}{1+\theta \hat q^2} - \frac{\theta \hat q^{\rho}\hat q_{\mu}}{1+\theta \hat q^2}\hat y_{\rho}].\label{7.8} 
\end{eqnarray}
The dilatation generator yielding the deformed transformations is given by,
\begin{equation}
\hat D = \frac{1}{2}[\hat y_{\rho}\frac{\hat q^{\rho}}{1+\theta \hat q^2} + \frac{\hat q^{\rho}}{1+\theta \hat q^2}\hat y_{\rho}]\label{7.9}
\end{equation}
so that 
\begin{eqnarray}
\delta \hat q_{\mu} &=& -i\epsilon[\hat q_{\mu}, \hat D]=-\epsilon \hat q_{\mu},\label{7.10}\\
\delta \hat y_{\mu} &=& -i\epsilon[\hat y_{\mu},\hat D]=\epsilon[\hat y_{\mu}-\hat y_{\rho}\frac{\theta \hat q^{\rho}\hat q_{\mu}}{1+\theta \hat q^2} - \frac{\theta \hat q^{\rho}\hat q_{\mu}}{1+\theta \hat q^2}\hat y_{\rho}].\label{7.11} 
\end{eqnarray}
In the limit $\theta\rightarrow {0}$, it reduces to the standard expression.
   The same result also follows on using the undeformed generators,
\begin{equation}
\hat D = \frac{1}{2}[\hat x^{\rho}\hat p_{\rho} +\hat p_{\rho}\hat x^{\rho}]\label{7.12}
\end{equation}  
and using the transformations (\ref{6.12}), (\ref{6.13}) which leads to the operator identity,
\begin{eqnarray}
\hat x^{\rho}\hat p_{\rho} +\hat p_{\rho}\hat x^{\rho} &=& \hat y_{\rho}\frac{\hat q^{\rho}}{1+\theta \hat q^2}+\frac{\hat q^{\rho}}{1+\theta \hat q^2}\hat y_{\rho}.\label{7.13}
\end{eqnarray}

Interpreted in this manner it is obvious that although $D$ in (\ref{7.9}) is deformed,
the corresponding algebra of generators remains the same
\begin{eqnarray}
&&\left[\hat D,\hat D\right]=0\label{7.14}\\
&&\left[\hat D,\hat q_{\mu}\right]=i \hat q_{\mu}\label{7.15}\\
&&\left[\hat D,\hat J_{\mu\nu}\right]=0.\label{7.16}
\end{eqnarray}
\subsection{Dynamical treatment}
 We next consider the dynamical model (\ref{3.7}) and study its classical invariance 
under the deformed dilatation transformation. We therefore take its 
massless version,
\begin{eqnarray}
S=\int d\tau[-\dot q^{\mu}y_{\mu}+\frac{\theta}{1+\theta q^2}(y.q)\dot q^{\mu}q_{\mu}-e q^2].
\label{7.2.1}
\end{eqnarray}
Since the demonstration of this invariance has certain distinctive features we
provide some computational details.

     The variations of the individual pieces turn out to be 
\begin{eqnarray}
&& -e\delta (q^2)  = 2\epsilon e q^2,\nonumber\\
&&\delta(-\dot q_{\mu}y^{\mu}) = \epsilon\frac{2\theta}{1+\theta q^2}(y\cdot q)(\dot q\cdot q),\nonumber\\
&&\delta(\frac{\theta}{1+\theta q^2}(y\cdot q)(\dot q\cdot q)) 
 = -\epsilon\frac{2\theta}{1+\theta q^2}(y\cdot q)(\dot q\cdot q).\nonumber
\end{eqnarray} 
Therefore the total variation in the Lagrangian is given by
\begin{eqnarray}
\delta L &=&2\epsilon e q^2. \label{7.2.2} 
\end{eqnarray}
This variation cannot be expressed as a total time derivative. However
if we pass to the constraint shell (\ref{3.18}) which in this case is given by
$q^2 = 0$, invariance is achieved, $\delta L = 0$. It is relevant
to note that had the mass term been included in the original Lagrangian,
its variation would still be given by (\ref{7.2.2}). In that case $q^2 =m^2 \ne 0$ 
and there is no invariance. This is compatible with the observation that
dilatation symmetry is broken for massive theories.
    
     We can reconstruct the dilatation generator by using Noether's theorem.
Using (\ref{5.5}), the variation (\ref{7.2}) and $B=0$ (since $\delta L=0$), we obtain,
\begin{eqnarray}
G &=& -\epsilon q^{\mu}(\frac{\theta (y\cdot q)q_{\mu}}{1+\theta q^2} - y_{\mu})
\nonumber\\
&=& \frac{\epsilon (y\cdot q)}{1+\theta q^2}.\label{7.2.3}
\end{eqnarray}

     It is possible to construct the operator analogue of the above generator by following the Weyl ordered prescription,
\begin{equation}
D = \frac{\epsilon}{4}[\hat y_{\rho}\frac{\hat q^{\rho}}{1+\theta \hat q^2} + \frac{\hat q^{\rho}}{1+\theta \hat q^2}\hat y_{\rho} + \hat q_{\rho}\hat y^{\rho}\frac{1}{1+\theta \hat q^2} +\frac{1}{1+\theta \hat q^2} \hat y^{\rho}\hat q_{\rho}].\label{7.2.4}
\end{equation}
The last two terms combine to give first two terms so that the final 
expression exactly agrees with (\ref{7.9}). 

     Further, to see whether the dilatation generator is gauge invariant
 or not, we calculate $\{D,q^2\}_{DB}$ and the result is
\begin{equation}
\{D,q^2\}_{DB} = 2q^2.\label{7.2.5}
\end{equation}  
Here also we see that right hand side of above equation is proportional
 to $q^2$. Therefore on the mass-shell constraint the Dirac bracket of $D$
with $q^2$ vanishes and hence we conclude that $D$ is a gauge invariant
 object. Hence, in spite of the deformation, the algebra of generators (\ref{7.14}), (\ref{7.15}) and (\ref{7.16})
remains the same. This was also inferred earlier from different considerations. Note that the algebra (\ref{7.2.5}) is also valid at the commutator level. From the basic algebra (\ref{2.1}) it can be shown that
\begin{equation}
[\hat D,\hat q^2] = 2i\hat q^2.
\end{equation}        
 
\subsection{Special Conformal Symmetry}
Here we discuss the deformed special conformal transformations. Since the 
computations are quite involved, we adopt a classical treatment that makes the
results transparent. Also, instead of proceeding from the basic requirement
of preserving the covariance of the Snyder algebra, we exploit the transformations (\ref{6.5}), (\ref{6.6}) to directly construct the deformed generator from the usual expression.
After getting the deformed generator, covariance of the Snyder algebra is
shown.  

      In the ordinary commutative space the generator for the special conformal transformation is given by,
\begin{equation}
K_{\mu} = 2x_{\mu}(x^{\rho}p_{\rho}) -x^2 p_{\mu}.\label{7.3.1}
\end{equation}
Using (\ref{6.5}), (\ref{6.6}) the deformed generator is computed from the above equation by a simple substitution,   
\begin{eqnarray}
K_{\mu}&=& -y^2q_{\mu} + 2y_{\mu}(y\cdot q) \frac{1}{1+\theta q^2}.\nonumber\\
&& + \theta ^2 (y\cdot q)^2 \frac{1}{(1+\theta q^2)^2}q^2 q_{\mu}.\label{7.3.2} 
\end{eqnarray}
The transformation rules for the deformed conformal transformation are given by 
\begin{eqnarray}
\begin{array}{rcl}
\delta y_{\mu} &=& \epsilon^{\nu}\{y_{\mu},K_{\nu}\}\\
\delta q_{\mu} &=& \epsilon^{\nu}\{q_{\mu},K_{\nu}\}
\end{array}
\label{7.3.3}
\end{eqnarray}
where $\epsilon^{\mu}$ is a constant infinitesimal parameter
corresponding to the special conformal transformation. Using the classical
Snyder algebra (\ref{3.23}) we obtain  
\begin{eqnarray}
\delta y_{\mu} &=& \{\theta y^2 q_{\mu}q_{\nu} - y^2 \delta_{\mu\nu} -6\theta
(y_{\nu}q_{\mu})(y\cdot q) \frac{1}{1+\theta q^2}\nonumber\\
&& + 2 y_{\mu}y_{\nu} + \theta ^2 (y\cdot q)^2 \frac{1}{(1+\theta q^2)^2}(q^2\delta_{\mu\nu}-\theta q_{\mu}q_{\nu}q^2 + 2q_{\mu}q_{\nu})\}\epsilon^{\nu}\label{7.3.4}\\
\delta q_{\mu} &=& \{2y_{\mu}q_{\nu} -2y_{\nu}q_{\mu} -2(y\cdot q) \frac{1}{1+\theta q^2}\delta_{\mu\nu}\}\epsilon^{\nu}.\label{7.3.5}
\end{eqnarray}
These are the deformed conformal transformations. Expectedly in the limit 
$\theta \rightarrow 0$ these reduce to the familiar structures in commutative
space. From these transformations it is found that 
\begin{eqnarray}
\begin{array}{rcl}
\delta\{y^{\mu},y^{\nu}\}_{DB} &=& \{\delta y^{\mu},y^{\nu}\} + \{y^{\mu},\delta y^{\mu}\}\\
&=&\theta\delta\left(y^{\mu} q^{\nu}- y^{\nu} q^{\mu}\right)\\
\delta\{q_{\mu},q_{\nu}\}_{DB}&=&0\\
\delta\{y^{\mu},q_{\nu}\}_{DB}&=&\delta\left(\delta^{\mu}_{ \nu}+\theta q^{\mu}q_{\nu}\right).
\end{array}
\label{7.3.6}
\end{eqnarray}
This is sufficient to prove that the compatibility of the deformed transformation with the Snyder brackets.
     Finally, we verify the invariance of the action (\ref{7.2.1}) under the deformed transformations. As shown earlier for the other generators, this also allows an alternative derivation of the conformal generator using Noether's theorem.  
        
 We first calculate the variation of the individual pieces in the action (\ref{7.2.1})
\begin{eqnarray}
-\delta({\dot q_{\mu}y^{\mu}})&=& \{\frac{d}{d\tau}(y^2 q_{\mu} - 2D y_{\mu}) + 
 2 \dot y_{\mu}\frac{y\cdot q}{1+\theta q^2} - 2\dot y_{\mu}\nonumber\\   
&& -\theta(\dot q\cdot q)K_{\mu} + \frac{d}{d\tau}(-2\theta y_{\mu}q^2 D + 
\theta^2 D^2 q^2 q_{\mu})\nonumber\\
&& + [\theta^2 D^2 \dot q_{\mu} - \frac{d}{d\tau}(-2\theta y_{\mu}D + \theta^2
D^2 q_{\mu})]q^2\}\epsilon^{\mu},\nonumber\\
\theta \delta (\frac{y\cdot q}{1+\theta q^2})(\dot q\cdot q) &=& -\{\theta K_{\mu} (\dot q\cdot q)\}\epsilon^{\mu},\nonumber\\
\theta (\frac{y\cdot q}{1+\theta q^2}) \delta (\dot q\cdot q) &=& \{\frac{d}{d\tau}(2D^2 \theta^2 q^2 q_{\mu}- 2Dy_{\mu}\theta q^2)- 2q^2(\theta^2q_{\mu} - 
 \theta  y_{\mu})\frac{dD}{d\tau}\}\epsilon^{\mu},\nonumber\\
-e\delta (q^2)&=& \{4 e q^2 (\theta D q_{\mu} - y_{\mu})\}\epsilon^{\mu}.\label{7.3.7}       
\end{eqnarray}
Some terms are expressible as a total time derivative which are retained since these will be useful in obtaining the generator. Terms not expressible in this way drop out for $q^2 = 0$. Combining all terms, we obtain,
\begin{eqnarray}
\delta L &=& \frac{d}{d\tau}[K_{\mu}\epsilon^{\mu}]\label{7.3.8}
\end{eqnarray} 
where $K_{\mu}$ is given in (\ref{7.3.2}). 

    As in the case of dilatation it is necessary to pass to the mass shell
constraint $q^2 = 0$ to get the invariance of the action. Of course for massive  theories $q^2 = m^2 \ne 0$ this invariance is broken.\\
\vspace{0.40cm}\\
{\bf Noether's theorem and generator} \ :\vspace{0.40cm}\\
The generator of deformed symmetry is alternatively computed from Noether's theorem using the definition (\ref{5.5}). The variation $\delta q^{\mu}$ is obtained from (\ref{7.3.5}) while B is abstracted from (\ref{7.3.8}). We find,
\begin{eqnarray}
G&=&2\epsilon^{\nu}(y_{\mu}q_{\nu} - y_{\nu}q_{\mu} -\frac{y\cdot q}{1+\theta q^2}
\delta_{\mu\nu})(\theta \frac{y\cdot q}{1+\theta q^2}q_{\mu} - y_{\mu})-\epsilon^{\nu}K_{\nu}\nonumber\\
&=&2\epsilon^{\nu}K_{\nu} -\epsilon^{\nu}K_{\nu}\nonumber\\
&=& \epsilon^{\nu}K_{\nu}\label{7.3.9}   
\end{eqnarray}     
 thereby reproducing the desired definition of the deformed generator given in (\ref{7.3.2})

      Further, we have verified that
\begin{eqnarray}
\{K_{\mu}, q^2\}_{DB}&=&(4\theta^2 q_{\mu}(y\cdot q)q^2 -4\theta(y\cdot q)q_{\mu} +2y_{\mu})q^2.\label{7.3.10} 
\end{eqnarray}  
Hence, on the mass-shell constraint $q^2=0$, $K_{\mu}$ is a gauge
invariant object.   

     The deformed generators obviously satisfy the complete (usual) conformal
algebra.  
\section{Snyder Algebra from Alternative Action}
The deformed symmetries preserving compatibility with the Snyder algebra were used to yield an action that possessed these symmetries. The symplectic structure of this first order action naturally yielded the Snyder algebra.
  In this section we propose an alternative form of the action which leads to the same algebra. This approach is more in tune with the conventional spirit but with important differences, where different actions\cite{rom,gir,ghosh,rome} have been suggested to yield the Snyder algebra. Usually the original action has a reparametrisation invariance which is eliminated by an appropriate choice of gauge. The Dirac brackets computed in the gauge fixed (reduced) space then correspond to (\ref{2.1}). In our approach, on the contrary, no gauge fixing is required and the cherished algebra follows naturally. The second point is that in usual approaches, the noncommutativity parameter $\theta$ is introduced by hand. In our analysis this parameter gets identified as $\theta\sim\frac{1}{m^2}$ where $m$ is the mass of a relativistic particle.

Our proposed action is given by,
\begin{eqnarray}
S &=& \int L(y,\dot y)d\tau\nonumber\\
&=&m\int d\tau\sqrt{g_{\mu\nu}\dot y^{\mu}\dot y^{\nu}}\label{8.1}
\end{eqnarray}
where
\begin{eqnarray}
&&\dot y^{\mu}=\frac{d y^{\mu}}{d \tau} \ {\textrm {and}}\label{8.2}\\
&&g_{\mu\nu}=\eta_{\mu\nu}-\frac{y_{\mu}y_{\nu}}{y^2}\label{8.3}
\end{eqnarray}

\begin{eqnarray}
y^2=y^{\sigma}y_{\sigma}=\eta_{\sigma\lambda}y^{\sigma}y^{\lambda}.\label{8.4}
\end{eqnarray}
Since the space-time coordinate $y^{\mu}(\tau)$ transforms as a scalar under reparametrisation
\begin{eqnarray}
&&\tau\rightarrow\tau'=\tau'(\tau)\label{8.8}\\
&&y^{\mu}(\tau)\rightarrow y'^{\mu}(\tau')=y^{\mu}(\tau)\label{8.9}
\end{eqnarray}
it is easy to verify that the system is reparametrisation invariant.

Note that $g_{\mu\nu}$ acts like a projection operator in the sense that
\begin{eqnarray}
&&g_{\mu\rho}g^{\mu}_{\nu}=g_{\rho\nu} \ (g^{\mu}_{\nu}=\eta^{\mu\rho}g_{\rho\nu})\label{8.5}\\
&&y^{\mu}g_{\mu\nu}=0.\label{8.6}
\end{eqnarray}
The canonical momentum is defined as,
\begin{eqnarray}
q_{\mu}&=&\frac{\partial L}{\partial \dot y^{\mu}}\nonumber\\
&=&m\frac{g_{\mu\nu}\dot y^{\nu}}{\sqrt{g_{\rho\sigma}\dot y^{\rho}\dot y^{\sigma}}}.\label{8.7}
\end{eqnarray}

Using (\ref{8.5}) and (\ref{8.6}) one immediately gets the following primary constraints,
\begin{eqnarray}
&&\phi_1=q^2-m^2\approx0\label{8.10}\\
&&\phi_2=q^{\mu}y_{\mu}\approx0.\label{8.11}
\end{eqnarray}
The first constraint is  the well known Einstein's relation for a relativistic particle of mass $m$. The second constraint is basically a transversality condition. It is interesting to note that this constraint does not involve the proper time so that reparametrisation invariance is kept intact. However the constraints do not close $(\{\phi_1,\phi_2\}=-2q^2=-2m^2)$ so that the symplectic structure is deformed. The deformed brackets are now computed by the standard Dirac procedure.

The constraint matrix is given by,
\begin{eqnarray}
\Lambda&=&\left(\matrix{\{\phi_1,\phi_1\}&\{\phi_1,\phi_2\}\cr\{\phi_2,\phi_1\}&\{\phi_2,\phi_2\} }\right)\nonumber\\
&=&\left(\matrix{0 & -2m^2 \cr 2m^2 & 0 }\right).\label{8.12}
\end{eqnarray}
Its inverse is
\begin{eqnarray}
\Lambda^{-1}=\left(\matrix{0 & \frac{1}{2m^2} \cr -\frac{1}{2m^2} & 0 }\right).\label{8.13}
\end{eqnarray}
Following (\ref{3.22}) the Dirac brackets are computed. They are given by,
\begin{eqnarray}
\begin{array}{rcl}
&&\{y_{\mu},y_{\nu}\}_{DB}=-\frac{1}{m^2}(y_{\mu}q_{\nu}-y_{\nu}q_{\mu})\\
&&\{y_{\mu},q_{\nu}\}_{DB}=\delta_{\mu\nu}-\frac{1}{m^2}q_{\mu}q_{\nu}\\
&&\{q_{\mu},q_{\nu}\}_{DB}=0.
\end{array}
\label{8.14}
\end{eqnarray}
This algebra reproduces the Snyder algebra  under the identification $\theta=-\frac{1}{m^2}$. Thus the relativistic action (\ref{8.1}) of mass $m=\sqrt{-\frac{1}{\theta}}$ basically describes the Snyder particle.

To show the internal consistency one can calculate the action from the constraints. The canonical Hamiltonian of the system,
\begin{eqnarray}
H_C=q_{\mu}\dot y^{\mu}-L=0\label{8.15}
\end{eqnarray}
vanishes, revealing its reparametrisation invariance. Hence the total Hamiltonian is just a linear combination of the primary constraints (\ref{8.10}) and (\ref{8.11}),
\begin{eqnarray}
H_T=\lambda_1(q^2 - m^2)+\lambda_{2}q^{\mu}y_{\mu}\label{8.16}
\end{eqnarray}
where $\lambda_1,\lambda_2$ are the Lagrange multipliers. The action is defined as,
 \begin{eqnarray}
S &=& \int d\tau\left(q^{\mu}\dot y_{\mu} -H_T\right)\label{8.17}\\
&=&\int d\tau \left(q^{\mu}\dot y_{\mu} -\lambda_1(q^2 - m^2)-\lambda_{2}q^{\mu}y_{\mu} \right). \label{8.18}
\end{eqnarray}
One can  eliminate $q_{\mu}$ from the above action by using its equation of motion,
\begin{eqnarray}
q_{\mu} &=& \frac{\dot y_{\mu} - \lambda_2 y_{\mu}}{2\lambda_1}.\label{8.19}
\end{eqnarray}
The equations of motion for $\lambda_1$ and $\lambda_2$ are,
\begin{eqnarray}\lambda_1 &=& \frac{1}{2m}\left[ \dot y^2 - \frac{(\dot y_{\mu}y^{\mu})^2}{y^2}\right],\label{8.20}\\
\lambda_2 &=& \frac{\dot y^{\mu}y_{\mu}}{y^2}. \label{8.21}  
\end{eqnarray}   
Substituting eq. (\ref{8.20}) and eq. (\ref{8.21}) in eq. (\ref{8.19}) and after doing some
 algebra we have 
\begin{eqnarray}
q_{\mu} = m\left(\dot y^2-\frac{(\dot y_{\mu}y^{\mu})^2}{y^2}\right)^{-\frac{1}{2}}
 \left( \dot y_{\mu} - \frac{\dot y^{\nu}y_{\nu}}{y^2}y_{\mu}\right). \label{8.22}
\end{eqnarray}
This is consistent with the result (\ref{8.7}).
    After substituting eq. (\ref{8.20}), eq. (\ref{8.21}) and eq. (\ref{8.22}) in eq. (\ref{8.18}) we have
\begin{equation}
S = m\int d\tau  \sqrt{\left[\dot y^2 - \frac{(\dot y^{\mu}y_{\mu})^2}{y^2}\right]}. \label{8.23}
\end{equation}
This action is precisely (\ref{8.1}).
This action is also invariant under the deformed Poincare transformations. For discussing the conformal transformations, the massless case has to be considered. In the form (\ref{8.23}) this is obviously not feasible. But this action is equivalently expressed in the form,
\begin{eqnarray}
S=\int d\tau[\frac{1}{4\lambda_1}(\dot y_{\mu}-\lambda_2y_{\mu})^2+\lambda_1m^2]\label{8.24}
\end{eqnarray}
where $\lambda_1,\lambda_2$ are given in (\ref{8.20}), (\ref{8.21}). Here the $m=0$ limit is easily implemented. Following the Dirac analysis a pair of primary constraints is obtained. From the definition of the canonical momenta,
\begin{eqnarray}
&&\pi_{\lambda_1}\approx0\label{8.25}\\
&&\pi_{\lambda_2}\approx0\label{8.26}\\
&&\pi_{\mu}-\frac{1}{2\lambda_1}(\dot y_{\mu}-\lambda_2y_{\mu})\approx0.\label{8.27}
\end{eqnarray}
The canonical Hamiltonian $H_C$ is given by
\begin{eqnarray}
H_C&=&\pi_{\mu}\dot y^{\mu}-L\label{8.28}\\
&=&\lambda_1q^2+\lambda_2y^{\mu}q_{\mu}\label{8.29}.
\end{eqnarray}
Time consistency of the primary constraints yields the secondary constraints,
\begin{eqnarray}
&&\{\pi_{\lambda_1},H_C\}=0\Rightarrow q^2\approx0\label{8.30}\\
&&\{\pi_{\lambda_2},H_C\}=0\Rightarrow y^{\mu}q_{\mu}\approx0.\label{8.31}
\end{eqnarray}
The interesting point is that these constraints (which are basically the $m=0$ value of (\ref{8.10}), (\ref{8.11})) are now first class. So the Poisson brackets remain valid. The Snyder algebra, which is obtained from the Dirac brackets, obviously does not arise here. Hence, for this action, invariance under deformations is not a meaningful issue.
 
\section{Conclusions}
Deformed conformal-Poincare symmetries compatible with Snyder algebra were
obtained in an algebraic approach. This approach is quite general to include
other types of noncommutative spaces. As an application we constructed dynamical models invariant under the deformed symmetries. From the dynamical content,
 a (classical)  mapping among the basic variables in the Snyder and usual (
commutative) descriptions was abstracted. A Weyl ordering of the classical map
 provided its quantum version. This was explicitly verified by comparing the relevant commutators. From this map a differential (momentum) representation
of the phase space operators in Snyder space was derived.

       An alternative action leading to the Snyder algebra was also given.
The new point here was that the noncommutativity parameter $\theta$ get
identified with the mass $m$ of a relativistic particle by 
$m=\sqrt{-\frac{1}{\theta}}$. The dispersion relation $p^2 =m^2$ remains valid but
 there is an inbuilt constraint enforcing a transversality condition that
deforms the basic (Poisson) algebra to the Snyder form.

     Perhaps a point worth mentioning is that, in deriving the Snyder algebra
from dynamical models, no gauge (or reparametrisation) fixing is required.
Since the basic variable ($y$) is not reparametrisation invariant, obtaining
the Snyder algebra in a specific gauge implies that it could be an artifact
 of the gauge. In this sense our derivation is conceptually clear than other
approaches\cite{rom,gir,ghosh,rome} where gauge fixing is mandatory. 

    As a future prospect we could construct field theory models with Snyder
 noncommutativity. An appropriate star product would have to be defined 
followed by demanding invariance under the deformed symmetries discussed here.      
\section*{Acknowledgment}
It is a pleasure to thank Subir Ghosh for discussions.    



\end{document}